\documentclass[10pt]{article}
\usepackage{latexsym}
\usepackage{amssymb}
\usepackage{amsmath}
\usepackage{amscd}
\usepackage{amsthm}
\usepackage[left=2cm,top=2.5cm,right=2.5cm,bottom=1.5cm]{geometry}
\usepackage[dvips]{graphicx}
\usepackage{hyperref}
\usepackage{textcomp}

\begin{document}
\begin{center}
\Large{\bf{An optimal system and invariant solutions of dark energy Models in cylindrically symmetric space-time}}

\vspace{10mm}

\normalsize{Anil Kumar Yadav$^{1}$ and Ahmad T Ali$^{2,3}$} \\

\vspace{4mm}
\normalsize{$^1$ Department of Physics, Galgotias College of Engineering and Technology,\\
 Knowledge Park - II, 
Greater Noida - 201306, India.\\
E-mail: abanilyadav@yahoo.co.in}\\
\vspace{2mm}
\normalsize{$^2$ King Abdul Aziz University,
Faculty of Science, Department of Mathematics,\\
PO Box 80203, Jeddah, 21589, Saudi Arabia.}\\
E-mail: atali71@yahoo.com\\
\normalsize{$^3$ Mathematics Department, Faculty of Science, Al-Azhar University,\\
Nasr city, 11884, Cairo, Egypt}\\
\end{center}

\begin{abstract}

In this paper, we derive some new invariant solutions of dark energy models in cylindrically symmetric
space-time. To quantify the deviation of pressure from isotropy, we introduce three different time
dependent skewness parameters along the spatial directions. The matter source consists of dark energy
which is minimally interact with perfect fluid. We use symmetry analysis method for solving the non-linear
partial differential equations (NLPDEs) which is more powerful than the classical methods of solving NLPDEs.
The geometrical and kinematical features of the models and the behaviour of the
anisotropy of dark energy, are examined in detail.

\end{abstract}



\section{Introduction }
The most striking discovery of the modern cosmology
is that the expansion of universe is accelerating at late time which is confirmed by
the high red-shift supernovae experiments (Riess et al \cite{riess1998}; Perlmutter et al \cite{perlmutter1999};
Bennet et al \cite{bennet2003};
Padmanabham \cite{pady2003}). Subsequent observations including more detailed studied of supernovae and
independent evidence from cluster of galaxies, LSS and CMBR confirmed and firmly established this
remarkable findings. The recent observations also suggest that dark energy is contributing about
73 $\%$ of the total energy of the universe (Spregel et al \cite{spregel2007}). There are various directions
aimed to construct the viable dark energy models such as scalar field models, dark fluid with complicated equation of
state parameter. However, the satisfactory explanation of dark energy origin is still unknown.

Many natural phenomena are described by a system of NLPDEs which is often difficult to be solved analytically,
since there is no general theory for completely solving NLPDEs. One
of the most useful techniques for finding exact solutions for the Einstein
field equations described by a system of NLPDEs is the symmetry method \cite{ali3, ali4}.
The existence of symmetries of differential equations under Lie group of
transformations often allows those equations to be reduced to simpler equations.
One of the major accomplishment of Lie was to identify that the properties
of global transformations of the group are completely and uniquely determined
by the infinitesimal transformations around the identity transformation.
This allows the non-linear relations for the identification of invariance groups to
be dealing with global transformation equations, we use differential operators,
called the group generators, whose exponentiation generates the action of the
group. The collection of these differential operators forms the basis for the Lie
algebra. There is a one-to-one correspondence between the Lie groups and the
associated Lie algebras. A basic problem concerning the group invariant solution is it's classification.
Since a Lie group usually contain infinitely many subgroups of the same dimensional, a classification of them up to
some equivalence relation is necessary. Ovsiannikov \cite{ovsi1} given equivalent of two
subalgebras of a given Lie algebra. Optimal system consists of representative
elements of each equality class. Discussion on optimal systems can be found in ref.
\cite{olve1}. Also Ibragimov \cite{ibra1}, in his paper
has given some examples of optimal system.

The study of universe on the scale in which anisotropy and inhomogeneity
are not ignored, cylindrically symmetric cosmological models play an important role. It
has a significant contribution in understanding some essential features of the universe such as the
formation of galaxies during the early stage of evolution of universe. Also the case of cylindrically
symmetry is natural because of mathematical simplicity of field equations whenever there exists a
direction in pressure and energy density are being equal. In the literature, Senovilla \cite{senovilla1990} obtained exact
solution of Einstein's field equation in cylindrically symmetric space-time. These are singularity free cosmological
model and satisfying energy and causality condition. Later on, Davinich et AL \cite{dadhich} have established
a link between the FRW model and the singularity free models. In our previous paper \cite{ali3, ali4},
we have also investigated the new class of exact solution of NLPDEs by symmetry group analysis method in
inhomogeneous space-time.

In this paper, we find one-dimensional optimal system for the Einstein field equations
for the dark energy model and classify reductions obtained by using one-dimensional
subalgebras. The paper is organized as follows: The metric and field equations are presented in section 2.
Section 3 and 4 deal with the symmetry analysis method and optimal system respectively.
The invariant solution of field equations are given in section 5. Section 6 deals with the physical and geometrical
properties of the models. Finally the results are discussed in section 7.\\

\section{The metric and field equations}

The space-time is given by
\begin{equation}
 \label{spacetime}
ds^{2}=A^{2}(dx^{2}-dt^{2})+B^{2}dy^{2}+C^{2}dz^{2}
\end{equation}
The metric potentials $A$, $B$ and $C$ are function of $x$ and $t$
The Einstien's field equation is given by
\begin{equation}
 \label{efe}
R^{i}_{j}-\frac{1}{2}g^{i}_{j}=-T^{(m)i}_{j}-T^{(de)i}_{j}
\end{equation}
where $T^{(m)i}_{j}$ and $T^{(de)i}_{j}$ are energy momentum tensor of perfect fluid and dark energy respectively.
These are given by\\
\begin{equation}
 \label{m}
T^{(m)i}_{j} = diag[-\rho^{(m)}, p^{(m)}, p^{(m)}, p^{(m)}]
\end{equation}
and
\begin{equation}
 \label{de}
T^{(de)i}_{j}=diag[-\rho^{(de)}, p^{(de)}_{x}, p^{(de)}_{y}, p^{(de)}_{z}] = diag[-1, \omega+\delta, \omega+\gamma,
\omega+\eta]\rho^{(de)}
\end{equation}
where $p^{(m)}$ and $\rho^{(de)}$ are, respectively the pressure and energy density of the perfect fluid
component; $\rho^{(de)}$ is the energy density of the DE components; $\delta(t)$, $\gamma(t)$ and $\eta(t)$
are skewness parameters along $x$, $y$ and $z$ axis respectively, which modify equation of state parameter of
dark energy.

In comoving coordinate system, the field equation (\ref{efe}), for the inhomogeneous
space-time (\ref{spacetime}), read as
\begin{equation}
 \label{efe1}
\frac{1}{A^{2}}\left[-\frac{\ddot{B}}{B}-\frac{\ddot{C}}{C}+\frac{\dot{A}}{A}\left(\frac{\dot{B}}{B}+\frac{\dot{C}}{C}\right)+
\frac{A^{\prime}}{A}\left(\frac{B^{\prime}}{B}+\frac{C^{\prime}}{C}\right)+\frac{B^{\prime}C^{\prime}}{BC}-
\frac{\dot{B}\dot{C}}{BC}\right]=p^{(m)}+(\omega + \delta)\rho^{(de)}
\end{equation}
\begin{equation}
 \label{efe2}
\frac{1}{A^{2}}\left[-\frac{\ddot{A}}{A}+\frac{\dot{A}^{2}}{A^{2}}+\frac{A^{\prime\prime}}{A}-
\frac{A^{{\prime}2}}{A^{2}}-\frac{\ddot{C}}{C}+\frac{C^{\prime\prime}}{C}\right]=p^{(m)}+(\omega + \gamma)\rho^{(de)}
\end{equation}
\begin{equation}
 \label{efe3}
\frac{1}{A^{2}}\left[-\frac{\ddot{A}}{A}+\frac{\dot{A}^{2}}{A^{2}}+\frac{A^{\prime\prime}}{A}-
\frac{A^{{\prime}2}}{A^{2}}-\frac{\ddot{B}}{B}+\frac{B^{\prime\prime}}{B}\right]=p^{(m)}+(\omega + \eta)\rho^{(de)}
\end{equation}
\begin{equation}
 \label{efe4}
\frac{1}{A^{2}}\left[-\frac{B^{\prime\prime}}{B}-\frac{C^{\prime\prime}}{C}+\frac{A^{\prime}}{A}\left(
\frac{B^{\prime}}{B}+\frac{C^{\prime}}{C}\right)+\frac{\dot{A}}{A}\left(\frac{\dot{B}}{B}+\frac{\dot{C}}{C}\right)
-\frac{B^{\prime}C^{\prime}}{BC}+\frac{\dot{B}\dot{C}}{BC}\right]=\rho^{(m)}+\rho^{(de)}
\end{equation}
\begin{equation}
 \label{efe5}
\frac{\dot{B}^{\prime}}{B}+\frac{\dot{C}^{\prime}}{C}-\frac{A^{\prime}}{A}\left(\frac{\dot{B}}{B}+\frac{\dot{C}}{C}\right)
-\frac{\dot{A}}{A}\left(\frac{B^{\prime}}{B}+\frac{C^{\prime}}{C}\right)=0
\end{equation}
Here $A^{\prime}=\frac{dA}{dx}$, $\dot{A} = \frac{dA}{dt}$ and so on.




The velocity field $u^i$ is ir-rotational. The scalar expansion
$\Theta$, shear scalar $\sigma^2$, acceleration vector $\dot{u}_i$
and proper volume $V$ are respectively found from the following
expressions \cite{dec1,rayc1}:

\begin{equation}  \label{u215}
\Theta\,=\,u_{;i}^{i}=\dfrac{1}{A}\Big(\dfrac{C_t}{C}+\dfrac{B_t}{B}+\dfrac{A_t}{A}\Big),
\end{equation}

\begin{equation}  \label{u216}
\begin{array}{ll}
\sigma^2\,=\,\dfrac{1}{2}\,\sigma_{ij}\,\sigma^{ij}=\dfrac{\Theta^2}{3}-\dfrac{1}{A^2}\Big(
\dfrac{B_t\,C_t}{B\,C}+\dfrac{A_t\,C_t}{A\,C}+\dfrac{A_t\,B_t}{A\,B}\Big),
\end{array}
\end{equation}

\begin{equation}  \label{u217}
\dot{u}_i\,=\,u_{i;j}\,u^j\,=\,\Big(\dfrac{A_x}{A},0,0,0\Big),
\end{equation}

\begin{equation}  \label{u218}
V=\sqrt{-g}=A^2\,B\,C,
\end{equation}
where $g$ is the determinant of the metric (\ref{spacetime}). The shear tensor is
\begin{equation}  \label{u219}
  \begin{array}{ll}
\sigma_{ij}\,=\,u_{(i;j)}+\dot{u}_{(i}\,u_{j)}-\frac{1}{3}\,\Theta\,(g_{ij}+u_i\,u_j).
\end{array}
\end{equation}
and the non-vanishing components of the $\sigma_i^j$ are
\begin{equation}  \label{u220}
\left\{
  \begin{array}{ll}
    \sigma_1^1\,&=\,\dfrac{1}{3\,A}\Big(\dfrac{2\,A_t}{A}-\dfrac{B_t}{B}-\dfrac{C_t}{C}\Big),\\
    \\
\sigma_2^2\,&=\,\dfrac{1}{3\,A}\Big(\dfrac{2\,B_t}{B}-\dfrac{C_t}{C}+\dfrac{A_t}{A}\Big),\\
\\
\sigma_3^3\,&=\,\dfrac{1}{3\,A}\Big(\dfrac{2\,C_t}{C}-\dfrac{B_t}{B}+\dfrac{A_t}{A}\Big),\\
\\
\sigma_4^4\,&=0.
   \end{array}
\right.
\end{equation}

The Einstein field equations (\ref{efe1})-(\ref{efe5}) constitute a system of five highly NLPDEs with six unknowns variables, $A$, $B$, $C$, $p^{(m)}$, $\rho^{(m)}$ and $\rho^{(de)}$. Therefore, one physically reasonable conditions amongst these parameters are required to obtain explicit solutions of the field equations. Let us assume that the expansion scalar $\Theta$ in the model (\ref{spacetime}) is proportional to the eigenvalue $\sigma_1^1$ of the shear tensor $\sigma_j^k$. Then from (\ref{u215}) and (\ref{u220}), we get
\begin{equation}\label{u223}
  \begin{array}{ll}
\dfrac{2\,A_t}{A}-\dfrac{B_t}{B}-\dfrac{C_t}{C}=3\,\gamma\,\Big(\dfrac{A_t}{A}+\dfrac{B_t}{B}+\dfrac{C_t}{C}\Big),
  \end{array}
\end{equation}
where $\gamma$ is a constant of proportionality. The above equation can be written in the form
\begin{equation}\label{u224}
  \begin{array}{ll}
\dfrac{A_t}{A}\,=n\,\Big(\dfrac{B_t}{B}+\dfrac{C_t}{C}\Big).
  \end{array}
\end{equation}
where $n=\dfrac{1+3\,\gamma}{2-3\,\gamma}$. If we integrate the above equation with respect to $t$, we can get the following relation
\begin{equation}\label{u225}
  \begin{array}{ll}
A(x,t)\,=\,f(x)\,\Big(B(x,t)\,C(x,t)\Big)^n,
  \end{array}
\end{equation}
where  $f(x)$ is a constant of integration which is an arbitrary function of $x$. If we substitute the metric function $A$ from (\ref{u217}) in the Einstein field equations, the equations (\ref{efe1})-(\ref{efe5}) transform to the NLPDEs of the coefficients $B$ and $C$ only, as the following new form:
\begin{equation}  \label{u210-1}
\begin{array}{ll}
    E_1=(\omega_y-\omega_z)\Bigg[\Big(\dfrac{f'}{f}\Big)'-\dfrac{f'}{f}\Big(\dfrac{B'}{B}+\dfrac{C'}{C}\Big)   -2\,n\Big(\dfrac{B^{\prime2}}{B^2}+\dfrac{C^{\prime2}}{C^2}\Big)-(2\,n+1)\dfrac{B'\,C'}{B\,C}-(2\,n-1)\dfrac{\dot{B}\,\dot{C}}{B\,C}\Bigg]\\
    \\
    \,\,\,\,\,\,\,\,\,\,\,\,\,\,\, \,\,\,\,\,\,\,\,\,\,\,\,\,\,\, \,\,\,\,\,\,\,\,\,\,\,\,\,\,\, \,\,\,\,\,\,\,\,\,\,\,\,\,\,\,
    +\Big(\omega_x-n\,\omega_y+(n-1)\omega_z\Big)\dfrac{\ddot{B}}{B}-\Big(\omega_x+(n-1)\omega_y-n\,\omega_z\Big)\dfrac{\ddot{C}}{C}\\
    \\
      \,\,\,\,\,\,\,\,\,\,\,\,\,\,\, \,\,\,\,\,\,\,\,\,\,\,\,\,\,\, \,\,\,\,\,\,\,\,\,\,\,\,\,\,\, \,\,\,\,\,\,\,\,\,\,\,\,\,\,\,
      -\Big(\omega_x-(n+1)\omega_y+n\,\omega_z\Big)\dfrac{B^{\prime\prime}}{B}+\Big(\omega_x+n\,\omega_y-(n+1)\omega_z\Big)\dfrac{C^{\prime\prime}}{C}
   =0,
  \end{array}
\end{equation}

\begin{equation}  \label{u210-2}
\begin{array}{ll}
        E_2=\dfrac{\dot{B}'}{B}+\dfrac{\dot{C}'}{C}-\Big(\dfrac{\dot{B}}{B}+\dfrac{\dot{C}}{C}\Big)
        \Bigg[2\,n\Big(\dfrac{B'}{B}+\dfrac{C'}{C}\Big)+\dfrac{f'}{f}\Bigg]=0,
  \end{array}
\end{equation}
where
\begin{equation}  \label{u210-3}
\begin{array}{ll}
        f^2\,A^2\,p^{(m)}(x,t)\,=\,\Big(\dfrac{f'}{f}\Big)'+n\,\Big(\dfrac{\dot{B}^2-B^{\prime2}}{B^2}+\dfrac{\dot{C}^2-C^{\prime2}}{C^2}\Big)
        -\big(\lambda\,\omega_y-n\big)\,\Big(\dfrac{\ddot{B}-B^{\prime\prime}}{B}\Big)\\
        \\
      \,\,\,\,\,\,\,\,\,\,\,\,\,\,\, \,\,\,\,\,\,\,\,\,\,\,\,\,\,\,\,\,\,\,
      \,\,\,\, \,\,\,\,\,\,\,\,\,\,\,\,\,\,\, \,\,\,\,\,\,\,\,\,\,\,\,\,\,\,
        -\big(n-\lambda\,\omega_z\big)\,\Big(\dfrac{\ddot{C}-C^{\prime\prime}}{C}\Big),
  \end{array}
\end{equation}

\begin{equation}  \label{u210-4}
\begin{array}{ll}
        f^2\,A^2\,\rho^{(m)}(x,t)\,=\,n\Big(\dfrac{\dot{B}^2+B^{\prime2}}{B^2}-\dfrac{\dot{C}^2+C^{\prime2}}{C^2}\Big)
        -\dfrac{f'}{f}\Big(\dfrac{B'}{B}+\dfrac{C'}{C}\Big)-(2\,n+1)\dfrac{\dot{B}\,\dot{C}}{B\,C}\\
        \\
      \,\,\,\,\,\,\,\,\,\,\,\,\,\,\, \,\,\,\,\,\,\,\,\,\,\,\,\,\,\,\,\,\,\,
      \,\,\,\, \,\,\,\,\,\,\,\,\,\,\,\,\,\,\, \,\,\,\,\,\,\,\,\,\,\,\,\,\,\,
        -(2\,n-1)\dfrac{B'\,C'}{B\,C}-\lambda\Big(\dfrac{\ddot{B}}{B}+\dfrac{\ddot{C}}{C}\Big)
        +(1+\lambda)\dfrac{B^{\prime\prime}}{B}+(1-\lambda)\dfrac{C^{\prime\prime}}{C},
  \end{array}
\end{equation}

\begin{equation}  \label{u210-5}
\begin{array}{ll}
        f^2\,A^2\,\rho^{(de)}(x,t)\,=\,\lambda\,\Big[\dfrac{\ddot{B}-B^{\prime\prime}}{B}+\dfrac{C^{\prime\prime}-\ddot{C}}{C}\Big],
  \end{array}
\end{equation}
and $\lambda=(\omega_y-\omega_z)^{-1}$.
\section{Symmetry analysis method}

In order to obtain an exact solutions of the Einstein field equations (\ref{u210-1})-(\ref{u210-5}), enough to
find a solution to the system of NLPDEs (\ref{u210-1})-(\ref{u210-2}).
The classical method for finding the solution is a separation method by
taking $B(x,t)=B_1(x)\,B_2(t)$ and $C(x,t)=C_1(x)\,C_2(t)$ \cite{ali09, pradhan2007, yadav2008}.
The symmetry
analysis method is a powerful method which gives an invariant solutions.  It is worth noting that: there are three major methods to
compute Lie point symmetries. The first one uses prolonged vector fields (Lie group method) which we will
use and explain it in details for this work. In the last century,
the application of this method has been developed by a number of mathematicians.
Ovsiannikov \cite{ovsi1}, Olver \cite{olve1}, Baumann \cite{Baumann00},
Bluman, G.W. and Anco \cite{Bluman02} are some of the mathematicians who have enormous amount of studies in this field.

The second utilizes differential forms (wedge products) due to Cartan
\cite{Cartan46}. A differential geometric approach to invariance groups and
solutions of PDEs was presented by Harrison and Estabrook \cite{Harrison71}.
They showed how to derive infinitesimal symmetries using Cartan's exterior
differential calculus. Edelen developed the theory of the differential form
method extensively and wrote some computer programs using this technique
\cite{Edelen81, Edelen85}. Suhubi \cite{Suhubi91} developed a differential
geometric method to find a set of explicit determination equations whose
solutions determine the components of the isovector fields. The isovector
fields, which are the infinitesimal generators of geometric transformations
with suitable algebraic invariance properties, are then used to obtain
invariant solutions of several PDEs which written in the balance form.
This method is applied for some equations such as: the heat equation \cite{Harrison97},
the generalized K-dV-Burger type equation \cite{Bhutani00}, the vacuum Maxwell equations \cite{Harrison05},
the Einstein vacuum equations \cite{ali1, ali2, att1},
the Biot's equations for one-dimensional linear poroelasticity \cite{OU07}.

The third one uses the notation of "formal symmetry"
\cite{Bocharov89, Bocharov90, Mikhailov90}. Although restricted
to evolution systems with two independent variables, this
method provides a very quick way to compute canonical
generalized symmetries. Due to its limited scope we will not
elaborate on that technique. For the first method, we write
\begin{equation}\label{u31}
\left\{
\begin{array}{ll}
x_i^{*}=x_i+\epsilon\,\xi_{i}(x_j,u_{\beta})+\bold{o}(\epsilon^2),\\
u_{\alpha}^{*}=u_{\alpha}+\epsilon\,\eta_{\alpha}(x_j,u_{\beta})+\bold{o}(\epsilon^2),
\end{array}
\right.
\,\,\,i,j,\alpha,\beta=1,2,
\end{equation}
as the infinitesimal Lie point transformations. We have assumed
that the system (\ref{u210-1})-(\ref{u210-2}) is invariant under the transformations given in
Eq. (\ref{u31}). The corresponding infinitesimal generator of Lie groups
(symmetries) is given by
\begin{equation}\label{u32}
 X=\sum_{i=1}^{2}\xi_{i}\dfrac{\partial}{\partial x_{i}}+\sum_{\alpha=1}^{2}\eta_{\alpha}
 \dfrac{\partial}{\partial u_{\alpha}},
 \end{equation}
where $x_1=x$, $x_2=t$, $u_1=B$ and $u_2=C$. The coefficients $\xi_{1}$, $\xi_{2}$, $\eta_{1}$ and $\eta_{2}$ are the functions of $x$, $t$, $B$ and $C$.
These coefficients are the components of infinitesimals symmetries
corresponding to $x$, $t$, $B$ and $C$ respectively, to be determined from the invariance conditions:
\begin{equation}\label{u33}
{\text{Pr}}^{(2)}\,X\Big(E_m\Big)|_{E_m=0}=0,
\end{equation}
where $E_m=0,\,m=1,2$ are the system (\ref{u210-1})-(\ref{u210-2}) under study and
${\text{Pr}}^{(2)}$ is the second prolongation of the symmetries $X$.
Since our equations (\ref{u210-1})-(\ref{u210-2}) are at most of order two, therefore, we
need second order prolongation of the infinitesimal generator
in Eq. (\ref{u33}). It is worth noting that, the $2$-th order prolongation is given by:
\begin{equation}\label{u34}
{\text{Pr}}^{(2)}\,X=\sum_{i=1}^{2}\xi_{i}\dfrac{\partial}{\partial x_{i}}+\sum_{\alpha=1}^{2}\eta_{\alpha}
 \dfrac{\partial}{\partial u_{\alpha}}+\sum_{i=1}^{2}\,\sum_{\alpha=1}^{2}\,\eta_{\alpha\,i}\,\dfrac{\partial}{\partial u_{\alpha,i}}
+\sum_{j=1}^{2}\,\sum_{i=1}^{2}\,\sum_{\alpha=1}^{2}\,\eta_{\alpha\,i\,j}\,\dfrac{\partial}{\partial u_{\alpha,ij}},
\end{equation}
where
\begin{equation}\label{u35}
\eta_{\alpha\,i}=D_{i}\Big(\eta_{\alpha}\Big)-\sum_{j=1}^{2}\,u_{\alpha,j}\,D_i\Big(\xi_j\Big)\,,\,\,\,\,\,\,\,
\eta_{\alpha\,i\,j}=D_{j}\Big(\eta_{\alpha\,i}\Big)-\sum_{k=1}^{2}\,u_{\alpha,k\,i}\,D_j\Big(\xi_k\Big)\,.
\end{equation}
The operator $D_{i}$ is called the {\it total derivative} ({\it Hach operator}) and taken the following
form:
\begin{equation}\label{u36}
D_i=\dfrac{\partial}{\partial x_i}+\sum_{\alpha=1}^{2}\,u_{\alpha,i}\,\dfrac{\partial}{\partial u_{\alpha}}
+\sum_{j=1}^{2}\,\sum_{\alpha=1}^{2}\,u_{\alpha,j\,i}\,
\dfrac{\partial}{\partial u_{\alpha,j}},
\end{equation}
where $u_{\alpha,i}=\frac{\partial u_{\alpha}}{\partial x_{i}}$ and $u_{\alpha,i\,j}=\frac{\partial^2 u_{\alpha}}{\partial x_{j}\,\partial x_{i}}$.

Expanding the system of Eqs. (\ref{u33}) along with the original system of Eqs. (\ref{u210-1})-(\ref{u210-2}) to eliminate $B_{xx}$ and $B_{xt}$ while we set the coefficients involving $C_{x}$, $C_{t}$, $C_{xx}$, $C_{xt}$, $C_{tt}$,
$B_{x}$, $B_{t}$, $B_{tt}$ and
various products to zero give rise the essential set of over-determined
equations. Solving the set of these determining equations, the components of symmetries takes the following form:
\begin{equation}\label{u37}
\xi_{1}=a_1\,x+a_2,\,\,\,\,\,\xi_{2}=a_1\,t+a_3,\,\,\,\,\,\eta_{1}=a_4\,B ,\,\,\,\,\,\eta_{2}=a_5\,C,
\end{equation}
such that the following conditions must be satisfied
\begin{equation}\label{u38} 
f(x)=a_6\,(a_1\,x+a_2)^{a_7}\,\,\,\,\,\,\,\,\,\,\omega_z=(1+a_8)\,\omega_x-a_8\,\omega_y,
\end{equation}
where $a_i,\,i=1,2,...,8$ are an arbitrary constants.

The characteristic equations associated to the general symmetries (\ref{u37}) are given by:
\begin{equation}\label{u41}
\dfrac{dx}{a_1\,x+a_2}=\dfrac{dt}{a_1\,t+a_3}=\dfrac{dB}{a_4\,B}=\dfrac{dC}{a_5\,C}.
\end{equation}

\section{Optimal system}
The general Lie point symmetries (\ref{u32}) becomes
\begin{equation}\label{u32-1}
 X=(a_1\,x+a_2)\dfrac{\partial}{\partial x}+(a_1\,t+a_3)\dfrac{\partial}{\partial t}
 +a_4\,B\,\dfrac{\partial}{\partial B}+a_5\,C\,\dfrac{\partial}{\partial C}.
 \end{equation}
Consequently, the non-linear Einstein field equations (\ref{u210-1})-(\ref{u210-2}) admits the 5-dimensional Lie algebra spanned by the independent symmetries shown below:
\begin{equation}\label{u32-2}
X_1=x\,\dfrac{\partial}{\partial x}+t\,\dfrac{\partial}{\partial t},\,\,\,\,X_2=\dfrac{\partial}{\partial x},\,\,\,\,
X_3=\dfrac{\partial}{\partial t},\,\,\,\,X_4=B\,\dfrac{\partial}{\partial B},\,\,\,\,X_5=C\,\dfrac{\partial}{\partial C}.
 \end{equation}
The forms of the symmetries $X_i$, $i=1,...,5$ suggest their significations: $X_2$, $X_3$ generate the symmetry of space translation, $X_1$, $X_4$, $X_5$ are associated with the scaling transformations. When the Lie algebra of these symmetries is computed, the only non-vanishing relations are:
\begin{equation}\label{u32-3}
[X_1,X_2]\,=-X_2,\,\,\,\,\,\,\,\,\,\,[X_1,X_3]=-X_3.
 \end{equation}

It is well known that reduction of the independent variables by one is possible using any linear combinations of the generators of symmetries (\ref{u32-2}). We will construct a set of minimal combinations known as optimal system \cite{olve1,ovsi1}. An optimal system of a Lie algebra is a set of $l$-dimensional subalgebra such that every $l$-dimensional is equivalent to a unique element of the set under some element of the adjoint representation. The adjoint representation of a Lie algebra $\{X_i,\,i=1,...,5\}$ is constructed using the formula:
\begin{equation}\label{u32-4}
\mathrm{Ad}(\exp[\varepsilon\,X_i])X_j=\sum_{k=0}^{\infty}\dfrac{\varepsilon^k}{k!}\Big(\mathrm{Ad(X_i)}\Big)^k\,X_j
=X_j-\varepsilon\,[X_i,X_j]+\dfrac{\varepsilon^2}{2}[X_i,[X_i,X_j]]-...\,.
 \end{equation}
In order to find the optimal system of the Einstein field equations (\ref{u210-1})-(\ref{u210-2}), first the following adjoint table is constituted as the following:
\begin{center}
\begin{tabular}{|c|c|c|c|c|c|}
  \multicolumn{6}{c}{}\\ \hline
  $\mathrm{Ad}$ & $X_1$ & $X_2$ & $X_3$ & $X_4$ & $X_5$  \\
\hline
  $X_1$ & $X_1$ & $\mathrm{e}^\varepsilon\,X_2$ & $\mathrm{e}^\varepsilon\,X_3$ & $X_4$ & $X_5$   \\
\hline
  $X_2$ & $X_1-\varepsilon\,X_2$ & $X_2$ & $X_3$ & $X_4$ & $X_5$  \\
\hline
  $X_3$ & $X_1-\varepsilon\,X_3$ & $X_2$ & $X_3$ & $X_4$ & $X_5$  \\
\hline
  $X_4$ & $X_1$ & $X_2$ & $X_3$ & $X_4$ & $X_5$  \\
\hline
  $X_5$ & $X_1$ & $X_2$ & $X_3$ & $X_4$ & $X_5$  \\
  \hline
\end{tabular}
\end{center}
Using simplification procedure in \cite{olve1,ovsi1}, we acquire an optimal system of one-dimensional subalgebras to be those spanned by:
\begin{equation}\label{u32-5}
\begin{array}{ll}
\{X^{(1)}=X_1+a_4\,X_4+a_5\,X_5,\,X^{(2)}=X_2+a_3\,X_3+a_4\,X_4+a_5\,X_5,\,\\
\,\,\,\,\,\,\,\,\,\,\,\,\,\,\,\,\,\,\,\,\,\,\,\,\,\,
X^{(3)}=X_3+a_4\,X_4+a_5\,X_5,\,X^{(4)}=X_4+a_5\,X_5,,\,X^{(5)}=,X_5\}.
\end{array}
\end{equation}

\section{Invariant solutions}

If we considered the symmetries $X^{(3)}$ or $X^{(4)}$ or $X^{(3)}$, then $a_1=a_2=0$. From equation (\ref{u38}) this leads to $f(x)=0$. So that, we shall analyse the invariant solutions associated with the optimal systems of symmetries $X^{(1)}$ and $X^{(2)}$ only as the following:\\

\textbf{Solution (I):} The symmetries $X^{(1)}$ has the characteristic equations:
\begin{equation}\label{u41-1}
\dfrac{dx}{x}=\dfrac{dt}{t}=\dfrac{dB}{a_4\,B}=\dfrac{dC}{a_5\,C}.
\end{equation}
Then the similarity variable and the similarity transformations takes the form:
\begin{equation}\label{u42-1}
\begin{array}{ll}
\xi=\dfrac{t}{x},\,\,\,\,\,\,B(x,t)=\,x^{a}\,\Psi(\xi),\,\,\,\,\,\,C(x,t)=\,x^{b}\,\Phi(\xi),
\end{array}
\end{equation}
where $a=a_4$ and $b=a_5$ are an arbitrary constants. In this case, we have
\begin{equation}\label{u42-2}
f(x)=c\,x^d,\,\,\,\,\,\,\,\,\,\,\omega_z=(1+q)\,\omega_x-q\,\omega_y,
\end{equation}
where $c=a_6$, $d=a_7$ and $q=a_8$ are an arbitrary constants. Substituting the transformations (\ref{u42-1}) in the field Eqs. (\ref{u210-1})-(\ref{u210-2}) lead
to the following system of ordinary differential equations:
\begin{equation}\label{u43-1}
\begin{array}{ll}
2\,n\,\xi^2\,\Big[\dfrac{\Psi'}{\Psi}+\dfrac{\Phi'}{\Phi}\Big]^2\,=\,\xi\,\Big[\dfrac{\Psi''}{\Psi}+\dfrac{\Phi''}{\Phi}\Big]+\dfrac{\alpha_a\,\Psi'}{\Psi}+\dfrac{\alpha_b\,\Phi'}{\Phi},
\end{array}
\end{equation}

\begin{equation}\label{u43-2}
\begin{array}{ll}
(1+q)\Bigg[2\,n\,\xi^2\,\Big(\dfrac{\Psi^{\prime2}}{\Psi^2}+\dfrac{\Phi^{\prime2}}{\Phi^2}\Big)
+\Big[\xi^2-1+2\,n\,(\xi^2-1)\Big]\dfrac{\Psi'\,\Phi'}{\Psi\,\Phi}\Bigg]+\alpha_3\,
=\,\xi\,\Big[\dfrac{\alpha_2\,\Psi'}{\Psi}+\dfrac{\alpha_1\,\Phi'}{\Phi}\Big]\\
\,\,\,\,\,\,\,\,\,\,\,\,\,\,\,\,\,\,\,\,\,\,\,\,\,
+\Big[q+\xi^2+n(1+q)(\xi^2-1)\Big]\dfrac{\Psi''}{\Psi}+\Big[1+q\,\xi^2+n(1+q)(\xi^2-1)\Big]\dfrac{\Phi''}{\Phi},
\end{array}
\end{equation}
where
\begin{equation}\label{u43-3}
\left\{
  \begin{array}{ll}
   \alpha_a=1+d+2\,n(a+b)-a,\\
\alpha_b=1+d+2\,n(a+b)-b,\\
\alpha_1=a+d+2\,n\,(1+a+b)+\Big[a+d+2(1-b)+2\,n(1+a+b)\Big]q,\\
\alpha_2=b+d+2n\,(1+a+b)+2(1-a)+\Big[b+d+2\,n\,(1+a+b)\Big]q,\\
\alpha_3=(1+b)(d+n\,b)+\Big[d+b\big(1+d+n+b(n-1)\big)\Big]q\\
\,\,\,\,\,\,\,\,\,\,\,\,\,\,\,\,\,\,\,\,
+a^2\big[n(1+q)-1\big]+a\Big[1+\big[d+n+b(2\,n+1)\big](1+q)\Big]
  \end{array}
\right.
\end{equation}
The equations (\ref{u43-1}) and (\ref{u43-2}) are non-linear ordinary differential equations (NLODEs) which is very difficult to solve. However, in a special cases, we can find a solution. Now, we suppose the following conditions:
\begin{equation}\label{u43-4}
\begin{array}{ll}
\dfrac{\Psi'}{\Psi}\,=\,\beta_1,\,\,\,\,\,\,\,\dfrac{\Phi'}{\Phi}\,=\,\beta_2.
\end{array}
\end{equation}
where $\beta_1$ and $\beta_2$ are an arbitrary constants. By integration the above equations, we get the following solution:
\begin{equation}\label{u43-6}
\begin{array}{ll}
\Psi(\xi)\,=\,\beta_3\,\exp\big[\beta_1\,\xi\big],\,\,\,\,\,\,\,\Phi(\xi)\,=\,\beta_4\,\exp\big[\beta_2\,\xi\big],
\end{array}
\end{equation}
where $\beta_3$ and $\beta_4$ are an arbitrary constants of integration. Substitute (\ref{u43-6}) in (\ref{u43-1}), we have the following condition:
\begin{equation}\label{u43-7}
\begin{array}{ll}
\Big[(2\,n-1)\big(\beta_1^2+\beta_2^2\big)+4\,n\,\beta_1\,\beta_2\Big]\xi=\alpha_a\,\beta_1+\alpha_b\,\beta_2.
\end{array}
\end{equation}
The coefficients of $\xi$ and the absolute value must be equal zero. Solving the two resulting conditions with respect to $n$ and $d$, we have:
\begin{equation}\label{u43-8}
\begin{array}{ll}
n=\dfrac{\beta_1^2+\beta_2^2}{2\big(\beta_1+\beta_2\big)^2},\,\,\,\,\,\,\,
d=-\dfrac{(1+a)\beta_1^2-(a+b-2)\beta_1\,\beta_2+(1+b)\beta_2^2}{\big(\beta_1+\beta_2\big)^2}.
\end{array}
\end{equation}
Substituting from (\ref{u43-8}) and  (\ref{u43-6}) into (\ref{u43-2}), we find the following condition:
\begin{equation}\label{u43-9}
\begin{array}{ll}
\beta_5\,\xi^2+\beta_6\,\xi+\beta_7\,=0,
\end{array}
\end{equation}
where
\begin{equation}\label{u43-10}
\left\{
  \begin{array}{ll}
   \beta_5=(\beta_1+\beta_2)^2\Big[(1-q)\big(\beta_1^2-\beta_2^2\big)-2(1+q)\beta_1\,\beta_2\Big],\\
\beta_6=2(\beta_1+\beta_2)\Big[\beta_1^2\big[2+b-a+q(a+b)\big]+2\beta_1\,\beta_2(b+a\,q)+\beta_2^2\big[b+a+q(2+a-b)\big]\Big],\\
\beta_7=\beta_2^2\Big[2+b(b-1)(q-1)+2q+a^2(q+3)+a\big[1+3q-2b(1+q)\big]\Big]\\
\,\,\,\,\,\,\,\,\,\,\,\,\,\,\,
+2\,\beta_1\,\beta_2\Big[(q-1)\big(b^2-b-a^2+a\big)+2(q+1)(1-2ab)\Big]\\
\,\,\,\,\,\,\,\,\,\,\,\,\,\,\,
+\beta_1^2\Big[2(1-q)(a-1)++2(q+1)(1-ab)+b\big[3+b+q(1+3b)\big]\Big]-\beta_5.
  \end{array}
\right.
\end{equation}
The condition (\ref{u43-9}) leads to $\beta_5\,=\,\beta_6\,=\,\beta_7\,=\,0$. Solving it with respect to $q$, $a$ and $b$, we have
\begin{equation}\label{u43-11}
\begin{array}{ll}
q=\dfrac{\beta_1^2-2\,\beta_1\beta_2-\beta_2^2}{\beta_1^2+2\,\beta_1\beta_2-\beta_2^2},\,\,\,\,\,
a=-\dfrac{(\beta_1+\beta_2)\big(\beta_1^3-3\,\beta_1^2\beta_2+\beta_1\beta_2^2+\beta_2^3\big)}{(\beta_1+\beta_2)^2},\,\,\,\,\,
b=\dfrac{(\beta_2^2-\beta_1^2)\big(\beta_1^2+2\,\beta_1\beta_2-\beta_2^2\big)}{(\beta_1+\beta_2)^2}.
\end{array}
\end{equation}
Now, by using (\ref{u43-11}) (\ref{u43-8}), (\ref{u43-6}), (\ref{u42-2}) and (\ref{u42-1}), we can find the solution if Einstein field equations as the following:
\begin{equation}\label{uu1}
\left\{
  \begin{array}{ll}
A(x,t)=\Big(\dfrac{\gamma_1}{x}\Big)\,\exp\Big[\dfrac{n\,(\beta_1+\beta_2)\,t}{x}\Big],\,\,\,\,\,\,\,\,\,\,
B(x,t)=\gamma_2\,x^{a}\,\exp\Big[\dfrac{\beta_1\,t}{x}\Big],\\
\\
C(x,t)=\gamma_3\,x^{b}\,\exp\Big[\dfrac{\beta_2\,t}{x}\Big],\,\,\,\,\,\,\,\,\,\,\,\,\,\,\,
\omega_z(t)=\dfrac{2\,\big(\beta_1^2-\beta_2^2\big)\,\omega_x(t)
-\big(\beta_1^2-2\beta_1\beta_2-\beta_2^2\big)\,\omega_y(t)}{\beta_1^2+2\beta_1\beta_2-\beta_2^2},
  \end{array}
\right.
\end{equation}
where $\gamma_1=c\,\Big(\beta_3\,\beta_4\Big)^n$, $\gamma_2=\beta_3$, $\gamma_3=\beta_4$, $\beta_1$ and $\beta_2$ are an arbitrary constants, while $\omega_x$ and $\omega_y$ are an arbitrary functions of $t$. It is observed from equations (\ref{uu1}), the line element (\ref{spacetime}) can be written in the following form:
\begin{equation}  \label{s1}
\begin{array}{ll}
ds_{1}^2=\Big(\dfrac{\gamma_1}{x}\Big)^2\,\exp\Big[\dfrac{2\,n\,(\beta_1+\beta_2)\,t}{x}\Big]\,\Big(dx^2-dt^2\Big)
+\gamma_2^2\,x^{2\,a}\,\exp\Big[\dfrac{2\,\beta_1\,t}{x}\Big]\,dy^2
+\gamma_3^2\,x^{2\,b}\,\exp\Big[\dfrac{2\,\beta_2\,t}{x}\Big]\,dz^2.
\end{array}
\end{equation}

\textbf{Remark:} In the above solution, we can replace $t$ by $t+\delta_1$ and $x$ by $x+\delta_2$ without loss of generality, where $\delta_1$ and $\delta_2$ are an some arbitrary constants.\\

\textbf{Solution (II):} The symmetries $X^{(2)}$ has the characteristic equations:
\begin{equation}\label{u51-1}
\dfrac{dx}{1}=\dfrac{dt}{a_3}=\dfrac{dB}{a_4\,B}=\dfrac{dC}{a_5\,C}.
\end{equation}
Then the similarity variable and the similarity transformations takes the form:
\begin{equation}\label{u52-1}
\begin{array}{ll}
\xi=a\,t-b\,x,\,\,\,\,\,\,B(x,t)=\Psi(\xi)\,\exp\big[c\,t\big],\,\,\,\,\,\,C(x,t)=\Phi(\xi)\,\exp\big[d\,t\big],
\end{array}
\end{equation}
where $a_3=\dfrac{b}{a}$, $c=\dfrac{a_4}{a_3}$ and $d=\dfrac{a_5}{a_3}$ are an arbitrary constants. In this case, we have
\begin{equation}\label{u52-2}
f(x)=p,\,\,\,\,\,\,\,\,\,\,\omega_z=(1+q)\,\omega_x-q\,\omega_y,
\end{equation}
where $p=a_6$ and $q=a_8$ are an arbitrary constants. Substituting the transformations (\ref{u52-1}) in the field Eqs. (\ref{u210-1})-(\ref{u210-2}), we can get the following system of ordinary differential equations:
\begin{equation}\label{u53-1}
\begin{array}{ll}
2\,n\,a\,\Big[\dfrac{\Psi'}{\Psi}+\dfrac{\Phi'}{\Phi}\Big]^2
+\dfrac{\alpha_c\,\Psi'}{\Psi}+\dfrac{\alpha_d\,\Phi'}{\Phi}\,=a\,\Big[\dfrac{\Psi''}{\Psi}+\dfrac{\Phi''}{\Phi}\Big],
\end{array}
\end{equation}

\begin{equation}\label{u53-2}
\begin{array}{ll}
\alpha_1+(1+q)\Bigg[2\,n\,b^2\,\Big(\dfrac{\Psi^{\prime2}}{\Psi^2}+\dfrac{\Phi^{\prime2}}{\Phi^2}\Big)
+\dfrac{\alpha_2\,\Psi'\,\Phi'}{\Psi\,\Phi}\Bigg]+\dfrac{\alpha_3\,\Psi'}{\Psi}+\dfrac{\alpha_4\,\Phi'}{\Phi}
+\dfrac{\alpha_a\,\Psi''}{\Psi}+\dfrac{\alpha_b\,\Phi''}{\Phi}=0,
\end{array}
\end{equation}
where
\begin{equation}\label{u53-3}
\left\{
  \begin{array}{ll}
\alpha_a=a^2\,\big[n\,(1+q)-1\big]-b^2\,\big[n\,(1+q)+q\big],\\
\alpha_b=a^2\,\big[n\,(1+q)-q\big]-b^2\,\big[n\,(1+q)+1\big],\\
\alpha_c=2\,n\,(c+d)-c,\\
\alpha_d=2\,n\,(c+d)-d,\\
\alpha_1=(c+d)\,\Big[c\,\big[n\,(1+q)-q\big]+d\,\big[n\,(1+q)-1\big]\Big],\\
\alpha_2=2\,n\,\big(a^2+b^2\big)+b^2-a^2,\\
\alpha_3=a\,\Big[(1+q)\,\alpha_d-2\,c\,q\Big],\\
\alpha_4=a\,\Big[(1+q)\,\alpha_c-2\,d\Big].
\end{array}
\right.
\end{equation}
The equations (\ref{u53-1}) and (\ref{u53-2}) are NLODEs which is very difficult to solve. However, in a special cases, we can find a solution. If we take $q\,=\,-1$, the equation (\ref{u53-2}) convert to the following simple form:
\begin{equation}\label{u53-4}
\begin{array}{ll}
\Big(\dfrac{a^2-b^2}{2\,a}\Big)\,\Big(\dfrac{\Psi''}{\Psi}-\dfrac{\Phi''}{\Phi}\Big)
+\dfrac{c\,\Psi'}{\Psi}-\dfrac{d\,\Phi'}{\Phi}+\dfrac{c^2-d^2}{2}=0.
\end{array}
\end{equation}
The above equation can be integrated when $b\,=\,\pm a$ and we get the following solution:
\begin{equation}\label{u53-5}
\begin{array}{ll}
\Phi(\xi)\,=\,\beta_1\,\Psi(\xi)\,\exp\Big[\Big(\dfrac{c^2-d^2}{2\,a\,d}\Big)\,\xi\Big],
\end{array}
\end{equation}
where $\beta_2$ is an arbitrary constants of integration. Substitute (\ref{u53-5}) in (\ref{u53-1}), we have the following equation:
\begin{equation}\label{u53-6}
\begin{array}{ll}
4\,a^2\,d\,(c+d)\dfrac{\Psi''}{\Psi}+4\,a\,c\,\Big[c^2+d^2-2\,n\,(c+d)^2\Big]\dfrac{\Psi'}{\Psi}\\
\,\,\,\,\,\,\,\,\,\,\,\,\,\,\,\,\,\,\,\,
+4\,a^2\,\Big[c^2-c\,d-2\,n\,(c+d)^2\Big]\dfrac{\Psi^{\prime2}}{\Psi^2}=2\,n\,(c+d)^2\,\big(c^2-d^2\big)+d^4-c^4.
\end{array}
\end{equation}
Under the transformation
\begin{equation}\label{u53-7}
\Psi'(\xi)\,=\,\Psi(\xi)\,\Omega(\xi)
\end{equation}
the above equation becomes:
\begin{equation}\label{u53-8}
\begin{array}{ll}
4\,a^2\,d\,(c+d)\,\Omega'=\Big[2\,n\,(c+d)^2-c^2-d^2\Big]\,\Big(c-d+2\,a\,\Omega\Big)\,\Big(c+d+2\,a\,\Omega\Big),
\end{array}
\end{equation}
By integrating the above equation, we have:
\begin{equation}\label{u53-9}
\begin{array}{ll}
\Omega(\xi)=\dfrac{d}{a-\beta_2\,\exp\Big[\Big(\dfrac{2\,n\,(c+d)^2-c^2-d^2}{a\,(c+d)}\Big)\xi\Big]}-\dfrac{c-d}{2\,a}.
\end{array}
\end{equation}
where $\beta_2$ is an arbitrary constant of integration. Substituting from (\ref{u53-9}) into (\ref{u53-7}) and integrating the resulting equation, we have:
\begin{equation}\label{u53-10}
\begin{array}{ll}
\Psi(\xi)=\beta_3\,\exp\Big[\Big(\dfrac{d-c}{2\,a}\Big)\xi\Big]\,
\Bigg(a-\beta_2\,\exp\Big[\Big(\dfrac{2\,n\,(c+d)^2-c^2-d^2}{a\,(c+d)}\Big)\xi\Big]\Bigg)^{\dfrac{d\,(c+d)}{c^2+d^2-2\,n\,(c+d)^2}},
\end{array}
\end{equation}
where $\beta_3$ is an arbitrary constant of integration. Therefore, from (\ref{u53-10}) and (\ref{u53-5}) we get:
\begin{equation}\label{u53-11}
\begin{array}{ll}
\Phi(\xi)=\beta_1\,\beta_3^{c/d}\,\exp\Big[\Big(\dfrac{c-d}{2\,a}\Big)\xi\Big]\,
\Bigg(a-\beta_2\,\exp\Big[\Big(\dfrac{2\,n\,(c+d)^2-c^2-d^2}{a\,(c+d)}\Big)\xi\Big]\Bigg)^{\dfrac{c\,(c+d)}{c^2+d^2-2\,n\,(c+d)^2}}.
\end{array}
\end{equation}
Now, by using (\ref{u53-11}), (\ref{u53-10}), (\ref{u52-2}) and (\ref{u52-1}), we can find the solution if Einstein field equations as the following:
\begin{equation}\label{uu2}
\left\{
  \begin{array}{ll}
A(x,t)=\gamma_1\,\exp\big[n\,(c+d)\,t\big]\,
\Bigg(a-\beta_2\,\exp\big[\gamma_4\,(t-x)\big]\Bigg)^{-\dfrac{n\,(c+d)}{\gamma_4}},\\
\\
B(x,t)=\gamma_2\,\exp\Big[\dfrac{(c+d)\,t+(c-d)\,x}{2}\Big]\,
\Bigg(a-\beta_2\,\exp\big[\gamma_4\,(t-x)\big]\Bigg)^{-\dfrac{d}{\gamma_4}},\\
\\
c(x,t)=\gamma_3\,\exp\Big[\dfrac{(c+d)\,t-(c-d)\,x}{2}\Big]\,
\Bigg(a-\beta_2\,\exp\big[\gamma_4\,(t-x)\big]\Bigg)^{-\dfrac{c}{\gamma_4}},\\
\\
\omega_z(t)=\omega_y(t),
  \end{array}
\right.
\end{equation}
where $\gamma_1=p\,\beta_1^n\,\beta_3^{n\,(c+d)/d}$, $\gamma_2=\beta_3$, $\gamma_3=\beta_1\,\beta_3^{c/d}$, $c$ $d$, $a$ $n$ and $\beta_2$ are an arbitrary constants, while $\omega_x$ and $\omega_y$ are an arbitrary functions of $t$ such that $\gamma_4=\dfrac{2\,n\,(c+d)^2-c^2-d^2}{c+d}$. It is observed from equations (\ref{uu2}), the line element (\ref{spacetime}) can be written in the following form:
\begin{equation}  \label{s2}
\begin{array}{ll}
ds_{2}^2=\gamma_1^2\,\exp\big[2\,n\,(c+d)\,t\big]\,
\Bigg(a-\beta_2\,\exp\big[\gamma_4\,(t-x)\big]\Bigg)^{-\dfrac{2\,n\,(c+d)}{\gamma_4}}\,(dx-dt)^2,\\
\\
\,\,\,\,\,\,\,\,\,\,\,\,\,\,\,\,\,\,\,\,\,\,\,\,\,\,\,\,\,\,
+\gamma_2^2\,\exp\big[(c+d)\,t+(c-d)\,x\big]\,
\Bigg(a-\beta_2\,\exp\big[\gamma_4\,(t-x)\big]\Bigg)^{-\dfrac{2\,d}{\gamma_4}}\,dy^2,\\
\\
\,\,\,\,\,\,\,\,\,\,\,\,\,\,\,\,\,\,\,\,\,\,\,\,\,\,\,\,\,\,
+\gamma_3^2\,\exp\big[(c+d)\,t-(c-d)\,x\big]\,
\Bigg(a-\beta_2\,\exp\big[\gamma_4\,(t-x)\big]\Bigg)^{-\dfrac{2\,c}{\gamma_4}}\,dz^2.
\end{array}
\end{equation}

\section{Physical and geometrical properties of the models}

\textbf{For the Model (\ref{s1}):}\\

The expressions of $p^{(m)}$, $\rho^{(m)}$ and $\rho^{(de)}$  for the model (\ref{s1}), are given by:

\begin{equation}\label{uu1-1}
  \begin{array}{ll}
p^{(m)}(x,t)=\dfrac{1}{\gamma_1^2\,x^2\,\big(\omega_x-\omega_y\big)}\,\Bigg[
\Big[\beta_2^2\,t^2+2\big[(1-b+n)\beta_2+n\,\beta_1\big]\,x\,t+\big[1+b(b-1)-\beta_2^2\big]\,x^2\Big]\,\omega_x\\
\,\,\,\,\,\,\,\,\,\,\,\,\,\,\,\,\,\,\,\,\,\,\,\,\,\,\,\,\,\,\,\,\,\,\,\,\,
-\Big[\big(\beta_1^2+q\,\beta_2^2\big)\,t^2+2\Big(\beta_1(1-a+n)+n\,\beta_2
+q\big[\beta_2(1-b+n)+n\,\beta_1\big]\Big)\,x\,t\\
\,\,\,\,\,\,\,\,\,\,\,\,\,\,\,\,\,\,\,\,\,\,\,\,\,\,\,\,\,\,\,\,\,\,\,\,\,
+\Big(1+a(a-1)-\beta_1^2+q\big[1+b(b-1)-\beta_2^2\big]\Big)\,x^2\Big]\,\Big(\dfrac{\omega_y}{(1+q)}\Big)
\Bigg]\,\exp\Big[-\dfrac{2\,n\,(\beta_1+\beta_2)\,t}{x}\Big],
  \end{array}
\end{equation}

\begin{equation}\label{uu1-2}
  \begin{array}{ll}
\rho^{(m)}(x,t)=\dfrac{1}{(1+q)\,\gamma_1^2\,x^2\,\big(\omega_x-\omega_y\big)}\,\Bigg[
(1+q)\,\big(\omega_x-\omega_y\big)\,\Big[\big[\big(\beta_1+\beta_2\big)^2-\beta_1\,\beta_2\big]\,t^2\\
\,\,\,\,\,\,\,\,\,\,\,\,\,\,\,\,\,\,\,\,\,\,\,\,\,\,\,\,\,\,\,\,\,\,\,\,\,\,\,\,\,\,\,\,\,\,\,
+\Big(\beta_1\big[1-b-2a+2n(a+b)\big]+\beta_2\big[1-2b-a+2n(a+b)\big]\Big)\,x\,t\\
\,\,\,\,\,\,\,\,\,\,\,\,\,\,\,\,\,\,\,\,\,\,\,\,\,\,\,\,\,\,\,\,\,\,\,\,\,\,\,\,\,\,\,\,\,\,\,
+\Big(a^2+ab+b^2-\beta_1\,\beta_2-n\,\big(\beta_1+\beta_2\big)^2\Big)\,x^2\Big]-G(x,y)
\Bigg]\,\exp\Big[-\dfrac{2\,n\,(\beta_1+\beta_2)\,t}{x}\Big],
  \end{array}
\end{equation}

\begin{equation}\label{uu1-3}
  \begin{array}{ll}
\rho^{(de)}(x,t)=\Bigg[\dfrac{G(x,t)}{(1+q)\,\gamma_1^2\,x^2\,\big(\omega_x-\omega_y\big)}\Bigg]\,\exp\Big[-\dfrac{2\,n\,(\beta_1+\beta_2)\,t}{x}\Big],
  \end{array}
\end{equation}
where
$$
G(x,t)=\big(\beta_1^2-\beta_2^2\big)\,t^2-2\big[(a-1)\beta_1+(1-b)\beta_2\big]\,x\,t
-\big[a-a^2-b+b^2+\beta_1^2-\beta_2^2\big]\,x^2.
$$
The volume element is
\begin{equation}  \label{uu1-4}
V=\gamma_1^2\,\gamma_2\,\gamma_3\,x^{a+b-2}\,\exp\Big[\dfrac{(2\,n+1)\,(\beta_1+\beta_2)\,t}{x}\Big].
\end{equation}
The expansion scalar, which determines the volume behavior of the fluid, is given by:
\begin{equation}\label{uu1-5}
  \begin{array}{ll}
\Theta=\Big(\dfrac{3\,\beta_1^2+4\,\beta_1\,\beta_2+3\,\beta_2^2}{2\,\gamma_1\,(\beta_1+\beta_2)}\Big)\,\exp\Big[-\dfrac{n\,(\beta_1+\beta_2)\,t}{x}\Big],
  \end{array}
\end{equation}
The non-vanishing components of the shear tensor, $\sigma_i^j$, are:
\begin{equation}\label{uu1-6}
  \begin{array}{ll}
\dfrac{\sigma_1^1}{\Theta}\,=\,-\dfrac{4\,\beta_1\,\beta_2}{9\,\beta_1^2+12\,\beta_1\,\beta_2+9\,\beta_2^2},
  \end{array}
\end{equation}

\begin{equation}\label{uu1-7}
  \begin{array}{ll}
\dfrac{\sigma_2^2}{\Theta}\,=\,\dfrac{3\,\beta_1^2+2\,\beta_1\,\beta_2-3\,\beta_2^2}{9\,\beta_1^2+12\,\beta_1\,\beta_2+9\,\beta_2^2},
  \end{array}
\end{equation}

\begin{equation}\label{uu1-8}
  \begin{array}{ll}
\dfrac{\sigma_3^3}{\Theta}\,=\,\dfrac{3\,\beta_2^2+2\,\beta_1\,\beta_2-3\,\beta_1^2}{9\,\beta_1^2+12\,\beta_1\,\beta_2+9\,\beta_2^2}.
  \end{array}
\end{equation}

The shear scalar is:
\begin{equation}\label{uu1-9}
  \begin{array}{ll}
\dfrac{\sigma^2}{\Theta^2}\,=\,\dfrac{3\,\beta_1^4-2\,\beta_1^2\,\beta_2^2+3\,\beta_2^4}{
3\,\big(3\,\beta_1^2+4\,\beta_1\,\beta_2+3\,\beta_2^2\big)^2}.
  \end{array}
\end{equation}

The acceleration vector is given by:
\begin{equation}\label{uu1-10}
  \begin{array}{ll}
\dot{u}_i=\Big[\dfrac{\big(\beta_1^2+\beta_2^2\big)\,t}{2\,(\beta_1+\beta_2)\,x^2}-\dfrac{1}{x}\Big]\Big(1,0,0,0\Big).
  \end{array}
\end{equation}
The deceleration parameter is given by \cite{dec1,rayc1}
\begin{equation}\label{uu1-11}
  \begin{array}{ll}
\mathbf{q}&=-3\,\Theta^2\,\Big(\Theta_{;i}\,u^{i}+\dfrac{1}{3}\,\Theta^2\Big)\\
&=-\Big(\dfrac{\beta_1\,\beta_1\,\big(3\,\beta_1^2+4\,\beta_1\,\beta_2
+3\,\beta_2^2\big)^3}{4\,\gamma_1^4\,\big(\beta_1+\beta_2\big)^4}\Big)\,\exp\Big[-\dfrac{4\,n\,(\beta_1+\beta_2)\,t}{x}\Big].
  \end{array}
\end{equation}

\textbf{For the Model (\ref{s2}):}\\

The expressions of $p^{(m)}$, $\rho^{(m)}$ and $\rho^{(de)}$  for the model (\ref{s2}), are given by:

\begin{equation}\label{uu2-1}
  \begin{array}{ll}
p^{(m)}(x,t)=\dfrac{\mathrm{e}^{-2\,n\,(c+d)\,t}}{\gamma_1^2\,\big(\omega_y-\omega_x\big)}\,
\Big[a-\beta_2\,\mathrm{e}^{\gamma_4\,(t-x)}\Big]^{2\,n\,(c+d)/\gamma_4-2}
\Bigg[c\,d\,\Big(a^2-\beta_2^2\,\mathrm{e}^{2\,\gamma_4\,(t-x)}\Big)\,\omega_x\\
\,\,\,\,\,\,\,\,\,\,\,\,\,\,\,\,\,\,\,\,\,\,\,\,\,\,\,\,
+\Big(a^2\,\big[(n-1)\,(c+d)^2+c\,d\big]-a\,\beta_2\,(c+d)\,\gamma_4\,\mathrm{e}^{\gamma_4\,(t-x)}
+\beta_2^2\,\big[n\,(c+d)^2+c\,d\big]\mathrm{e}^{2\,\gamma_4\,(t-x)}\Big)\,\omega_y\Bigg],
  \end{array}
\end{equation}

\begin{equation}\label{uu2-2}
  \begin{array}{ll}
\rho^{(m)}(x,t)=\dfrac{\mathrm{e}^{-2\,n\,(c+d)\,t}}{\gamma_1^2}\,
\Big[a-\beta_2\,\mathrm{e}^{\gamma_4\,(t-x)}\Big]^{2\,n\,(c+d)/\gamma_4-2}
\Bigg[\dfrac{1}{\omega_y-\omega_x}\Bigg(a^2\,\big[n\,(c+d)^2-c^2-d^2\big]\\
\,\,\,\,\,\,\,\,\,\,\,\,\,\,\,\,\,\,\,\,\,\,\,\,\,\,\,\,
-a\,\beta_2\,(c+d)\,\gamma_4\,\mathrm{e}^{\gamma_4\,(t-x)}+n\,\beta_2^2\,(c+d)^2\,\mathrm{e}^{2\,\gamma_4\,(t-x)}\Bigg)
-a^2\,\big[n\,(c+d)^2+c\,d\big]\\
\,\,\,\,\,\,\,\,\,\,\,\,\,\,\,\,\,\,\,\,\,\,\,\,\,\,\,\,
+a\,\beta_2\,(c+d)\,\gamma_4\,\mathrm{e}^{\gamma_4\,(t-x)}
-\beta_2^2\,\big[(n-1)\,(d+c)^2+c\,d\big]\,\mathrm{e}^{2\,\gamma_4\,(t-x)}\Bigg],
  \end{array}
\end{equation}

\begin{equation}\label{uu2-3}
  \begin{array}{ll}
\rho^{(de)}(x,t)=\dfrac{\mathrm{e}^{-2\,n\,(c+d)\,t}}{\gamma_1^2\,(\omega_x-\omega_y)}\,
\Big[a-\beta_2\,\mathrm{e}^{\gamma_4\,(t-x)}\Big]^{2\,n\,(c+d)/\gamma_4-2}
\Big[a^2\,\big[n\,(c+d)^2-c^2-d^2\big]\\
\,\,\,\,\,\,\,\,\,\,\,\,\,\,\,\,\,\,\,\,\,\,\,\,\,\,\,\,\,\,\,\,\,\,\,\,\,\,
-a\,\beta_2\,(c+d)\,\gamma_4\,\mathrm{e}^{\gamma_4\,(t-x)}+n\,\beta_2^2\,(c+d)^2\,\mathrm{e}^{2\,\gamma_4\,(t-x)}\Big].
  \end{array}
\end{equation}

The volume element is
\begin{equation}  \label{uu2-4}
V=\gamma_1^2\,\gamma_2\,\gamma_3\,\mathrm{e}^{(2\,n+1)\,(c+d)\,t}\,
\Big[a-\beta_2\,\mathrm{e}^{\gamma_4\,(t-x)}\Big]^{-(2\,n+1)\,(c+d)/\gamma_4}.
\end{equation}
The expansion scalar, which determines the volume behavior of the fluid, is given by:
\begin{equation}\label{uu2-5}
  \begin{array}{ll}
\Theta=\dfrac{a\,(n+1)\,(c+d)\,\mathrm{e}^{-n\,(c+d)\,t}}{\gamma_1^2}\,
\Big[a-\beta_2\,\mathrm{e}^{\gamma_4\,(t-x)}\Big]^{n\,(c+d)/\gamma_4-1},
  \end{array}
\end{equation}
The non-vanishing components of the shear tensor, $\sigma_i^j$, are:
\begin{equation}\label{uu2-6}
  \begin{array}{ll}
\dfrac{\sigma_1^1}{\Theta}\,=\,\dfrac{2\,n-1}{3\,(n+1)},
  \end{array}
\end{equation}

\begin{equation}\label{uu2-7}
  \begin{array}{ll}
\dfrac{\sigma_2^2}{\Theta}\,=\,\dfrac{a\,(1-2\,n)\,(c+d)-3\,\beta_2^2\,(c-d)\,\mathrm{e}^{\gamma_4\,(t-x)}}{6\,a\,(n+1)\,(c+d)},
  \end{array}
\end{equation}

\begin{equation}\label{uu2-8}
  \begin{array}{ll}
\dfrac{\sigma_3^3}{\Theta}\,=\,\dfrac{a\,(1-2\,n)\,(c+d)+3\,\beta_2^2\,(c-d)\,\mathrm{e}^{\gamma_4\,(t-x)}}{6\,a\,(n+1)\,(c+d)}.
  \end{array}
\end{equation}

The shear scalar is:
\begin{equation}\label{uu2-9}
  \begin{array}{ll}
\dfrac{\sigma^2}{\Theta^2}\,=\,\dfrac{1}{12}\Big(\dfrac{1-2\,n}{1+n}\Big)^2
+\Bigg(\dfrac{\beta_2\,(c-d)\,\mathrm{e}^{\gamma_4\,(t-x)}}{2\,a\,(n+1)\,(c+d)}\Bigg)^2.
  \end{array}
\end{equation}

The acceleration vector is given by:
\begin{equation}\label{uu2-10}
  \begin{array}{ll}
\dot{u}_i=\dfrac{n\,\beta_2\,(c+d)}{\beta_2-a\,\mathrm{e}^{\gamma_4\,(x-t)}}\,\Big(1,0,0,0\Big).
  \end{array}
\end{equation}
The deceleration parameter is given by:
\begin{equation}\label{uu2-11}
  \begin{array}{ll}
\mathbf{q}=\Big(\dfrac{a^3\,(1+n)^3\,(c+d)^3}{\gamma_1^4}\Big)\,
\mathrm{e}^{-4\,n\,(c+d)\,t}\,\Big[a\,(2\,n-1)\,(c+d)-3\,\beta_2\,\gamma_4\,\mathrm{e}^{\gamma_4\,(x-t)}\Big]\,
\Big[a-\beta_2\,\mathrm{e}^{\gamma_4\,(x-t)}\Big]^{4\,n\,(c+d)/\gamma_4-4}.
  \end{array}
\end{equation}

\section {Conclusion}
In this paper, we have investigated an optimal system and invariant solutions of dark energy models in cylindrically
symmetric space-time. Generally, the models represent shearing, expanding and non-rotating universe in which
flow vector is geodetic. On the basis of optimal systems of symmetries $X^{(1)}$ and $X^{(2)}$, we obtained
two models (\ref{s1}) and (\ref{s2}) respectively. For model (\ref{s1}), our study reveals:\\
\begin{itemize}
 \item $\dfrac{\sigma}{\theta}$ = constant forever, therefore the model does not approach isotropy.
\item The declaration parameter is negative, therefore it represents the model of accelerating universe.
\item As $t \rightarrow \infty$, $V \rightarrow \infty$ but $\rho^{(de)} \rightarrow 0$ hence volume increases
with grow of time while dark energy density decreases.
\end{itemize}
For model(\ref{s2}) , we have following observations\\
\begin{itemize}
 \item $\lim_{t \rightarrow 0}\dfrac{\sigma}{\theta}$ =  constant.
\item The deceleration parameter is positive hence it represents the model of decelerating universe.
\end{itemize}
Thus, in our analysis, model (\ref{s1}) is most suitable model of universe matches with observations.


\end{document}